\documentstyle[preprint,pra,aps]{revtex}

\begin{document}

\bibliographystyle{prsty}
\draft
\tighten

\title{Information and noise in quantum measurement}


\author{Holger F. Hofmann\\Department of Physics, Faculty of Science, 
University of Tokyo\\7-3-1 Hongo, Bunkyo-ku, Tokyo113-0033, Japan}

\date{\today}

\maketitle

\begin{abstract}
Even though measurement results obtained in the real world are generally
both noisy and continuous, quantum measurement
theory tends to emphasize the ideal limit of perfect precision and
quantized measurement results. In this article, a more general concept
of noisy measurements is applied to investigate the role of quantum noise
in the measurement process. 
In particular, it is shown that the effects
of quantum noise can be separated from the effects of information obtained
in the measurement. However, quantum noise is required to ``cover up''
negative probabilities arising as the quantum limit is approached.
These negative probabilities represent fundamental quantum mechanical 
correlations between the measured variable and the variables affected by
quantum noise.  
\end{abstract}

\section{Introduction}

The interpretation of quantum mechanical measurements and the connection
between the classical regime and the quantum regime still remains somewhat
of a mystery, even after nearly a century of quantum physics\cite{Whe83}. 
When introducing quantum mechanics, most textbooks and university courses 
tend to throw out
all classical physics and start with mathematical axioms instead. Although
no one would doubt that the concepts of classical physics are still successful
in describing most of our everyday experience, there seems to be no gradual
approach which introduces quantum modifications to an otherwise unaltered
classical world. In fact, recent controversies \cite{PT99} suggest that 
many physicists
are meanwhile prepared to consider the classical limit as a somewhat special
case justified only as a crude approximation or even as a subjective illusion
(e.g. in the popular many-worlds-interpretation)\cite{Mer96}. 
At the heart of these 
controversies is the problem of ``measurement''. Classical theories were
able to ignore the problem, because one could always assume knowledge of
all facts. In quantum mechanics, there is a disturbing separation between
the continuous deterministic evolution of a system state and the random
selection of quantized results in the measurement. 

The practical problems of interpreting the physical meaning of quantum
states are solved by the statistical interpretation provided through the
measurement postulate. However, no such postulate exists in classical physics.
What would then be the analogy between quantum measurement and classical 
measurement? In the early days of quantum physics, such considerations 
resulted in the formulation
of the well known uncertainty principle \cite{Hei27}.
All measurements must introduce some noise into the system in order
to preserve the uncertainty relations. If the precision of a measurement 
approaches the quantum limit, the noise introduced during the measurement
process completely obscures the original values of all physical properties
which are not eigenvalues of the observed state. If the precision of the 
measurement is far below the quantum limit, the noise introduced 
into the measured system may be negligibly small and the classical 
situation is reproduced in most respects. It should be noted, however, 
that the assumption
of infinitely precise coordinates can not be recovered - classical physics
is the physics of low precision and noisy observations, just as 
non-relativistic physics is the physics of low velocities. The 
mathematical representation of classical coordinates in terms of real numbers 
is therefore an approximation of the quantum mechanical reality of finite
precision. 

In order to illustrate the transition from the classical (noisy) world
to the world of quantization, it seems to be desirable to include 
the precision of measurements in the measurement theory. Indeed, several 
discussions of measurements with limited precision have been presented,
usually based on the standard measurement postulate and an intermediate 
system.
Much of the focus has been on continuous measurements, which can 
be described by quantum trajectories \cite{Car93,Doh99,Hof98}. 
Some rather remarkable
properties obtained by combining weak (i.e. low precision) measurements
with standard projective measurements have been pointed out by Aharonov 
and coworkers\cite{Aha88,Aha93}. 
In the following, the concept of weak measurement is taken 
one step further by the introduction of a generalized measurement postulate
for both weak and strong measurements. This eliminates the requirement
of distinguishing qualitatively between two situations and allows a more
direct approach to the transition from classical low precision measurements
to quantum mechanical precision. Moreover, the generalized measurement
postulate can be applied directly to the quantum system considered, 
without the introduction of a measurement system. 

The generalized measurement postulate can be interpreted entirely in terms
of classical measurement processes. Therefore, the noise introduced into
the system during the measurement interaction can be separated from the
information obtained.  
If the precision of the measurement is sufficiently low to avoid a resolution
of quantization, this separation allows a classical interpretation
of the information obtained in the measurement. 
As precision increases, however, the noise effect is required to ``cover
up'' negative probabilities predicted by the information-induced change
in the system state. These negative probabilities can then be interpreted as 
nonclassical correlations between the measurement result and the quantum
fluctuations in the observed system. 

\section{The generalized measurement postulate}

\subsection{Limitations of precise measurements}
The axiomatic introduction to quantum mechanics given e.g. in von Neumanns
``Mathematical Foundations of Quantum Mechanics''\cite{Neu55} 
usually emphasizes the 
beauty and simplicity of the mathematical structure. It is therefore not
surprising that the measurement postulate does not include the possibility 
of uncertainty in the measurement result which is emphasized by Heisenberg 
in his ``Physical Principles of Quantum Theory''\cite{Hei58}. 
Instead, von Neumann merely shows that 
the measurement postulate based on an infinitely precise observation of 
eigenvalues is consistent with equally precise indirect measurement 
performed by letting the system interact with a meter. However, 
the uncertainty relations require that 
infinitely precise knowledge of one variable can only be obtained
by introducing an infinite amount of uncertainty in another variable.
A precise measurement of position always requires an infinite uncertainty 
of momentum, a precise measurement of the intensity
of a radio signal would require complete uncertainty of the phase, 
and a precise measurement of the angular momentum of the moon would 
require a complete delocalization of the moon itself.
A large number of similar examples can be constructed, showing that
any quantum mechanically precise measurement on a macroscopic system 
would cause macroscopic uncertainties. Thus, the projective measurement
postulate has no classical limit and consequently fails to  
describe some of the most typical classical measurements performed 
on macroscopic objects.

In order to amend this shortcoming and to illustrate the continuous
transition from quantum mechanics to classical 
physics, it is therefore desirable to replace the abstract mathematical
definition of measurement given by von Neumann with a formulation
closer to everyday experience. Actually, this can be achieved without 
any change in the fundamental structure of quantum theory, since the
choice of a projection on eigenstates was not motivated by 
physical observations, but rather by considerations of mathematical 
simplicity. 
The generalized formalism may appear to be less elegant, but it
faithfully reproduces all physical results, including a more natural 
transition to the classical limit.

\subsection{The generalized measurement operator}
If a careful experimentalist
obtains the result that a variable A corresponding to a hermitian operator
$\hat{A}$ in quantum theory has a value of $\bar{A}\pm \delta\! A$, then 
we need not assume that the experimentalist performed a projective
measurement on A and failed to read out the correct result. Instead,
the uncertainty of $\delta\! A$ may be a consequence of the experimentalist's
attempts to minimize the noise introduced into the system according to the
uncertainty principle. Typical examples of such measurements are optical 
back action evasion quantum nondemolition measurements of photon number 
\cite{Lev86,Fri92} and of quadrature components of the light field 
\cite{Por89,Per94}. 
In all these experimental realizations, the measurement resolution was
finite and the coherence between eigenstates was reduced but not lost
due to the measurement interaction \cite{Imo85,Kit87}.

In order to obtain a measurement value in a backaction evasion measurement, 
the observed system must be coupled
to a measurement apparatus. Ideally, a meter variable $\hat{x}$ of the 
measurement device is shifted by an amount equal to the system variable
$\hat{A}$. This interaction is described by the unitary
operator $\hat{S}$ which transforms the eigenstates $\mid x; A \rangle$ of
the pointer variable $\hat{x}$ and the system variable $\hat{A}$ according to
\begin{equation}
\hat{S}\mid x;A\rangle = \mid x+A; A \rangle.
\end{equation}
The shift in $\hat{x}$ may also be expressed using the conjugate
meter variable $\hat{p}$, which is defined by the operator property
\begin{equation}
\hat{x}\hat{p}-\hat{p}\hat{x} = i.
\end{equation}
The interaction operator then reads
\begin{equation}
\hat{S} = \exp\left(-i \hat{p} \hat{A} \right).
\end{equation}
The meter variable $\hat{p}$ thus acts on all system variables which
do not commute with $\hat{A}$, causing noise in the system. In order
to perform a measurement of $\hat{A}$ with a resolution of $\delta\!A$,
it is necessary to prepare the initial meter state in a Gaussian
wavepacket with a variance of $\delta\!A^2$ in $\hat{x}$. According to the
uncertainty principle, this requires a variance of $1/(2 \delta\!A)^2$ 
in $\hat{p}$. In order to obtain 
information about the system, it is necessary to read out only $\hat{x}$,
thereby eliminating all information about the noise variable $\hat{p}$.
  
For an initial Gaussian meter state and an arbitrary system state 
$\mid \Phi_S \rangle$, the entangled state of meter and 
system after the measurement interaction reads
\begin{equation}
\mid \psi_f \rangle =
\int d\!x \sum_A 
(2\pi \delta\! A^2)^{-1/4}\; 
              \exp \left(-\frac{(A-x)^2}{4 \delta\! A^2}\right)
\langle A \mid \Phi_S \rangle \; \mid x; A \rangle
.
\end{equation}
A measurement readout of the meter variable $\hat{x}$ selects a subspace
of the total Hilbert space. Within this subspace, the system state 
corresponding to a measurement readout of $\mid x=\bar{A} \rangle$ is given by
\begin{equation}
\hat{P}_{\delta\!A} (\bar{A}) \mid \Phi_S \rangle
=\sum_A 
(2\pi \delta\! A^2)^{-1/4}\; 
              \exp \left(-\frac{(A-\bar{A})^2}{4 \delta\! A^2}\right)
\langle A \mid \Phi_S \rangle \; \mid A \rangle.
\end{equation}
The effect of a measurement of $\hat{A}$ with finite resolution $\delta\!A$
is therefore described by a set of generalized measurement operators 
$\hat{P}_{\delta A}(\bar{A})$ corresponding to the possible measurement 
results $\bar{A}$, 
\begin{equation}
\hat{P}_{\delta A}(\bar{A})= (2\pi \delta\! A^2)^{-1/4}\; 
              \exp \left(-\frac{(\hat{A}-\bar{A})^2}{4 \delta\! A^2}\right)
.
\end{equation}
Instead of projecting the system state into an eigenstate of
$\hat{A}$, this operator modifies the statistical weight of each
eigenstate, while preserving as much coherence as possible \cite{Hel76}.

The probability distribution $p(\bar{A})$ over possible measurement 
results $\bar{A}$ corresponding to a given initial state 
$\mid \psi_{i} \rangle$ is given by
\begin{eqnarray}
\label{eq:distrib}
p(\bar{A}) &=& 
\langle \psi_i \mid \hat{P}^\dagger_{\delta A}(\bar{A})
                    \hat{P}_{\delta A}(\bar{A})\mid \psi_{i} \rangle
\nonumber \\ &=&
\langle \psi_i \mid \left(\hat{P}_{\delta A}(\bar{A})\right)^2
                    \mid \psi_{i} \rangle
.
\end{eqnarray}
This probability distribution may be characterized by the averages
$\langle \bar{A} \rangle$ and $\langle \bar{A}^2 \rangle$. These
averages are related to the quantum mechanical expectation values by
\begin{eqnarray}
\langle \bar{A} \rangle &=& \langle \hat{A} \rangle
\nonumber \\
\langle \bar{A}^2 \rangle &=& \langle \hat{A}^2 \rangle + \delta\! A^2.
\end{eqnarray}
Thus the average results of the general measurement postulate correspond
to the expectation value and the total variance is equal to the sum of the
variance in the system and the squared uncertainty of the measurement.

It is important to understand that the additional fluctuations in the 
measurement result do not correspond to additional noise in the system.
Instead, increasing the noise in the measurement result decreases the 
noise introduced in the system according to the uncertainty principle.
After the measurement, the system state will have changed to
\begin{equation}
\mid \psi_{f}(\bar{A}) \rangle = \frac{1}{\sqrt{p(\bar{A})}}
\hat{P}_{\delta A}(\bar{A})\mid \psi_{i} \rangle.
\end{equation}
This is still a pure state. The extent to which it differs from the
initial state is determined both by the decomposition of the 
initial state $\mid \psi_{i} \rangle$ in terms of eigenstates of
$\hat{A}$, and by the uncertainty $\delta\! A^2$ of the measurement.
For very large uncertainties, the changes in the system state
are only weak, illustrating the low level of noise in the measurement
interaction.

For infinitely precise measurements,
the properties of the generalized measurement operator correspond
to the properties of the projection operator. The projective 
measurement postulate is thus reproduced by $\delta\! A\to 0$. 
Since most actual measurements have a finite resolution, however,
the generalized measurement operator $\hat{P}_{\delta A}(\bar{A})$
provides a more realistic description of measurements than the
original projection postulate. Discussions of quantum mechanical
phenomena may be simplified considerably by accepting the validity 
of the generalized measurement postulate without requiring the 
formal derivation given above. In particular, the relationship between
the measurement information obtained and the noise introduced in the 
measurement can be investigated in far greater detail, directly revealing
fundamental properties of the operator formalism in the measurement
data. Note that in the following, the generalized measurement postulate 
will be applied without further reference to the actual physical 
interaction by which
the measurement result $\bar{A}$ is obtained. The measurement resolution
$\delta A$ is considered to be a property characterizing the measurement
effect on the system rather than a property of the external measurement
setup. It is then possible to focus on the system properties as the source 
of the measurement statistics. 

\subsection{Density matrix formulation}
It is a straightforward matter to apply the generalized measurement
postulate to an initial mixed state density matrix $\hat{\rho}_i$. 
The probability and the effect of the measurement result $\bar{A}$ 
on the density matrix read
\begin{eqnarray}
\label{eq:project}
p(\bar{A}) &=& 
Tr\left\{ \hat{P}_{\delta A}(\bar{A}) \hat{\rho}_i
                    \hat{P}^\dagger_{\delta A}(\bar{A}) \right \}
\nonumber \\ &=&
Tr\left\{ \left(\hat{P}_{\delta A}(\bar{A})\right)^2
                    \hat{\rho}_i \right \}
\nonumber \\
\hat{\rho_{f}}(\bar{A}) &=& \frac{1}{p(\bar{A})}
\hat{P}_{\delta A}(\bar{A}) \hat{\rho}_i
                    \hat{P}^\dagger_{\delta A}(\bar{A})
.
\end{eqnarray}
Using the density matrix formulation, the physical effect of the measurement
interaction on the system may be determined by mixing the final state 
density matrices $\hat{\rho_{f}(\bar{A})}$ according to their respective 
statistical weights $p(\bar{A})$,
\begin{eqnarray}
\label{eq:total}
\hat{\rho}_f (\mbox{total}) &=& \int d\bar{A}\; p(\bar{A}) \hat{\rho}_{f}(\bar{A})
\nonumber \\ &=&
\int d\bar{A}\; \hat{P}_{\delta\! A}(\bar{A}) \hat{\rho}_i
              \hat{P}^\dagger_{\delta\! A}(\bar{A}).
\end{eqnarray}
This density matrix represents
the system between the measurement interaction and the readout of the
measurement result. It therefore describes the (average) noise effect 
caused by the measurement interaction. In particular, the elements of the
density matrix which describe coherence between eigenstates of $\hat{A}$
are reduced by 
\begin{equation}
\label{eq:element}
\langle A_1\mid\hat{\rho}_f (\mbox{total})\mid A_2 \rangle =    
                \exp\left(-\frac{(A_1-A_2)^2}{8 \delta\!A^2}\right)
         \langle A_1\mid \hat{\rho}_i \mid A_2 \rangle
\end{equation}
where $\mid A_1 \rangle$ and $\mid A_2 \rangle$ are eigenstates of 
$\hat{A}$ with eigenvalues of $A_1$ and $A_2$, respectively. 
Thus there is a gradual decrease of coherence depending only on the
separation of the eigenvalues $A_1$ and $A_2$. The Gaussian dependence
of the suppression factor on the difference of the eigenvalues indicates
that the decoherence effect is extremely sensitive to the relationship
between the eigenvalue difference $|A_1-A_2|$ and the resolution $\delta\! A$
of the measurement. Indeed, the decoherence factor is greater than 0.88 for 
$|A_1-A_2| < \delta\! A$ and lower than 0.14 for $|A_1-A_2| > 4 \delta\! A$. 
This rapid transition from almost no decoherence to almost complete 
decoherence corresponds to the notion that the ability to distinguish the
eigenvalues $A_1$ and $A_2$ requires decoherence between the corresponding
eigenstates. If the separation of eigenvalues $|A_1 - A_2|$ is large,
even a very weak measurement which otherwise preserves microscopic 
coherences will destroy the coherence between the eigenstates of 
$A_1$ and $A_2$. It is therefore much more difficult to preserve the
quantum coherence between states with quantitatively different physical
properties than to preserve coherence between quantitatively similar 
states\cite{Har98}.

Since equation (\ref{eq:total}) makes no reference to the measurement
result actually obtained, it represents only the effects of physical
interaction involved in the measurement. Thus it is equivalent to a
description of decoherence in open systems interacting with an unknown
environment. However, in the case of a quantum measurement, the 
meter takes the place of the environment, and the meter information
is recovered in the measurement. Consequently, it is not possible to
average over the meter state and interpretational problems related to 
the entanglement of system and meter can arise as soon as the actual 
information obtained in the measurement is considered. Specifically,
the measurement readout requires an interpretation of 
$\hat{\rho}_f (\mbox{total})$ as a mixture corresponding to the 
different possible measurement results, while the more simple alternative
of assuming random phase noise in the coherence between eigenstates of 
$\hat{A}$ cannot be recovered. 

\section{Separation of information and noise}
\subsection{Formal separation}
The total effects of a measurement result of $\bar{A}\pm\delta\! A$
on the density matrix element 
$\langle A_1\mid\hat{\rho}_i \mid A_2 \rangle$ is given by equation 
(\ref{eq:project}). In terms of matrix elements of the eigenstates of
$\hat{A}$ it reads
\begin{eqnarray}
\label{eq:readout}
\lefteqn{
\langle A_1\mid\hat{\rho}_f (\bar{A})\mid A_2 \rangle =} 
\nonumber \\ &&
\frac{1}{\sqrt{2\pi \delta\! A^2}\; p(\bar{A})} 
\exp\left(-\frac{\left(\frac{A_1+A_2}{2}-\bar{A}\right)^2}
                {2 \delta\!A^2}\right)\!
\exp\left(-\frac{(A_1-A_2)^2}{8 \delta\!A^2}\right)
\langle A_1\mid \hat{\rho}_i \mid A_2 \rangle
\nonumber \\ &=& \frac{1}{\sqrt{2\pi \delta\! A^2}\; p(\bar{A})}
   \exp\left(-\frac{\left(\frac{A_1+A_2}{2}-\bar{A}\right)^2}
                   {2 \delta\!A^2}\right)
          \langle A_1\mid\hat{\rho}_f (\mbox{total})\mid A_2 \rangle 
.
\end{eqnarray}
Since the effect of decoherence given in equation (\ref{eq:total})
can be identified with the Gaussian factor changing the matrix
element of the initial density matrix $\hat{\rho}_i$ to  
the corresponding matrix element of $\hat{\rho}_f(\mbox{total})$,
the remaining factor should describe the effect of a noise free measurement.
This factorization of the decoherence factor and the factor 
associated with the measurement result obtained allows an
unambiguous separation of information and noise in quantum
measurements, even though these two aspects are connected by the
requirements of the uncertainty relations. It is therefore possible
to overcome the uncertainty limitations and to examine the structure
of quantum mechanical reality which is hidden beneath the noise.

The measurement may be interpreted as a two step process. 
In the first step, decoherence is caused by the physical
interaction between the system and the measurement setup, changing the 
density matrix from $\hat{\rho}_i$ to $\hat{\rho}_f(\mbox{total})$.
In the second step, information about the system is obtained without 
any (additional) interaction. This step may actually occur far away from 
the system. While the first step involves well defined physical processes,
the second step relates to a change in the probabilistic expression of
the system state due to information gained about the system.
Equation (\ref{eq:total}) shows that the total density matrix
$\hat{\rho}_f(\mbox{total})$ can be interpreted as a mixture of all possible 
final density matrices $\hat{\rho}_f(\bar{A})$, so it is possible to
consider the change from
$\hat{\rho}_f(\mbox{total})$ to $\hat{\rho}_f(\bar{A})$ as a selection of
a reality which existed before the information was obtained. It therefore
seems that the classical separation of information and physical interaction
has been preserved. However, 
the properties of $\hat{\rho}_f(\mbox{total})$ have been modified by
the decoherence in step one, and it is impossible to remove this step
without violating the uncertainty principle. 
In particular, the entanglement 
between system and meter ensures that the noise introduced into the system
can never be compensated once a measurement readout is obtained. 

\subsection{Simulation of noise free measurements}
The only difference between the classical measurement situation and the
quantum mechanical situation is the uncertainty in the measurement 
interaction. Except for the required relationship between decoherence
and measurement resolution, the two measurement steps can be separated.
In a classical situation, the noise added in the measurement interaction
is both undesirable and avoidable. It is assumed that the information
obtained in the measurement refers to a reality of A which exists as
an element of reality regardless of the measurement. By examining the
quantum mechanical version of a noise free measurement, it is possible to
find out ``what is wrong with these classical elements of reality''
\cite{Mer90}.

The procedure for selecting a sub-ensemble density matrix 
$\hat{\rho}_f(\bar{A})$ from the total density matrix  
$\hat{\rho}(\mbox{total})$ described by equation (\ref{eq:total})
can be applied directly to the initial density matrix $\hat{\rho}_i$.
It is then possible to reverse the actual sequence of steps
in the measurement process in order to investigate the changes in 
the system state caused by the information obtained before quantum 
noise is added. 
In a noise free measurement, the initial density matrix $\hat{\rho}_i$
is decomposed in analogy with equation (\ref{eq:total}),
\begin{equation}
\label{eq:decomp}
\hat{\rho}_i = \int d\bar{A}\; p(\bar{A}) \hat{\rho}_{m}(\bar{A}).
\end{equation}
The matrix elements of the density matrix $\hat{\rho}_m(\bar{A})$ 
describing the effects of measurement without noise then read
\begin{eqnarray}
\label{eq:midrho}
\lefteqn{
\langle A_1\mid\hat{\rho}_m (\bar{A})\mid A_2 \rangle =} 
\nonumber \\ &&
 \frac{1}{\sqrt{2\pi \delta\! A^2}\; p(\bar{A})}
   \exp\left(-\frac{\left(\frac{1}{2}(A_1+A_2)-\bar{A}\right)^2}
                   {2 \delta\!A^2}\right)
          \langle A_1\mid\hat{\rho}_i \mid A_2 \rangle.
\end{eqnarray}
Indeed, the statistical weight factor modifying the
density matrix looks harmless enough. The matrix element is enhanced
or suppressed depending on the closeness
of the average quantum number $(A_1+A_2)/2$ of the matrix element 
to the measurement result $\bar{A}$. This effect would correspond to the
classically expected modification if $(A_1+A_2)/2$ would somehow represent 
the (classical) value of $\hat{A}$ associated with the matrix element.
However, the matrix element indicates a coherent superposition of two 
different eigenvalues $A_1$ and $A_2$ of $\hat{A}$. Therefore the modification
of its statistical weight should be represented by separate contributions
from $A_1$ and $A_2$. Since this is not so, however, a serious problem 
arises concerning the relation
\begin{equation}
\label{eq:positivity}
 \langle A_1\mid\hat{\rho}_m \mid A_2 \rangle
 \langle A_2\mid\hat{\rho}_m \mid A_1 \rangle
\leq
 \langle A_1\mid\hat{\rho}_m \mid A_1 \rangle
 \langle A_2\mid\hat{\rho}_m \mid A_2 \rangle,
\end{equation}
which guarantees that all probabilities obtained as expectation values 
of the density matrix $\hat{\rho}_m$ are positive. Condition 
(\ref{eq:positivity}) is only fulfilled if
\begin{equation}
\label{eq:condition}
 \frac{\langle A_1\mid\hat{\rho}_i \mid A_2 \rangle
       \langle A_2\mid\hat{\rho}_i \mid A_1 \rangle}
      {\langle A_1\mid\hat{\rho}_i \mid A_1 \rangle
       \langle A_2\mid\hat{\rho}_i \mid A_2 \rangle}
\leq
\exp\left(-\frac{(A_1-A_2)^2}{4\delta\! A^2}\right)
.
\end{equation}
Thus the decoherence factor of equations (\ref{eq:element}) and 
(\ref{eq:readout}) reappears in a requirement which can only be
fulfilled if the coherence of the density matrix $\hat{\rho}_i$
is sufficiently low. 
If the coherence of $\hat{\rho}_i$ is high, however, condition
(\ref{eq:positivity}) is violated and consequently negative probabilities
are obtained for some of the possible coherent superpositions
of the eigenstates $\mid A_1 \rangle$ and $\mid A_2 \rangle$.

The difference between quantum mechanics and classical physics thus emerges
as the measurement information obtained at low resolution is not only
information on eigenvalues of $\hat{A}$, but also on the average values
of off-diagonal matrix elements. Negative probabilities arise naturally, 
because there may be no possible eigenvalues corresponding to $(A_1+A_2)/2$.
Nevertheless, measurement results close to  $(A_1+A_2)/2$ indicate, that 
the coherent contribution of the corresponding off-diagonal matrix element
is greater than the associated diagonal elements. In classical physics, 
the reality of $A$ would be well defined. In the quantum formalism, however,
the eigenvalues of $\hat{A}$ represent only an incomplete description
of the reality of the operator variable $\hat{A}$. 

\subsection{Illustration of negative probabilities in a two level system}
\label{sec:example}
\begin{figure}
 
\begin{picture}(380,250)
\put(0,120){\vector(1,0){360}}
\put(360,100){\makebox(20,20){$\langle\hat{s}_x\rangle$}}
\put(180,10){\vector(0,1){220}}
\put(190,230){\makebox(20,20){$\langle\hat{s}_z\rangle$}}
\put(160,100){\makebox(20,20){0}}
\put(80,110){\line(0,1){20}}
\put(60,100){\makebox(20,20){$-\frac{1}{2}$}}
\put(280,110){\line(0,1){20}}
\put(260,100){\makebox(20,20){$\frac{1}{2}$}}
\put(170,220){\line(1,0){20}}
\put(160,200){\makebox(20,20){$\frac{1}{2}$}}
\put(170,20){\line(1,0){20}}
\put(160,0){\makebox(20,20){$-\frac{1}{2}$}}
\bezier{200}(180,220)(222,220)(251,191)
\bezier{200}(251,191)(280,162)(280,120)
\bezier{200}(280,120)(280,78)(251,49)
\bezier{200}(251,49)(222,20)(180,20)
\bezier{200}(180,20)(138,20)(109,49)
\bezier{200}(109,49)(80,78)(80,120)
\bezier{200}(80,120)(80,162)(109,191)
\bezier{200}(109,191)(138,220)(180,220)
\bezier{200}(180,220)(249,220)(298,191)
\bezier{200}(298,191)(347,162)(347,120)
\bezier{200}(347,120)(347,78)(298,49)
\bezier{200}(298,49)(249,20)(180,20)
\bezier{200}(180,20)(111,20)(62,49)
\bezier{200}(62,49)(13,78)(13,120)
\bezier{200}(13,120)(13,162)(62,191)
\bezier{200}(62,191)(111,220)(180,220)
\end{picture}

\caption{\label{bloch} Illustration of the expectation values of the 
spin-1/2 system after the noise free measurement (ellipse) and after the 
complete measurement (circle) for a measurement uncertainty of $\delta\! 
s_z = 1/2$. }
\end{figure}
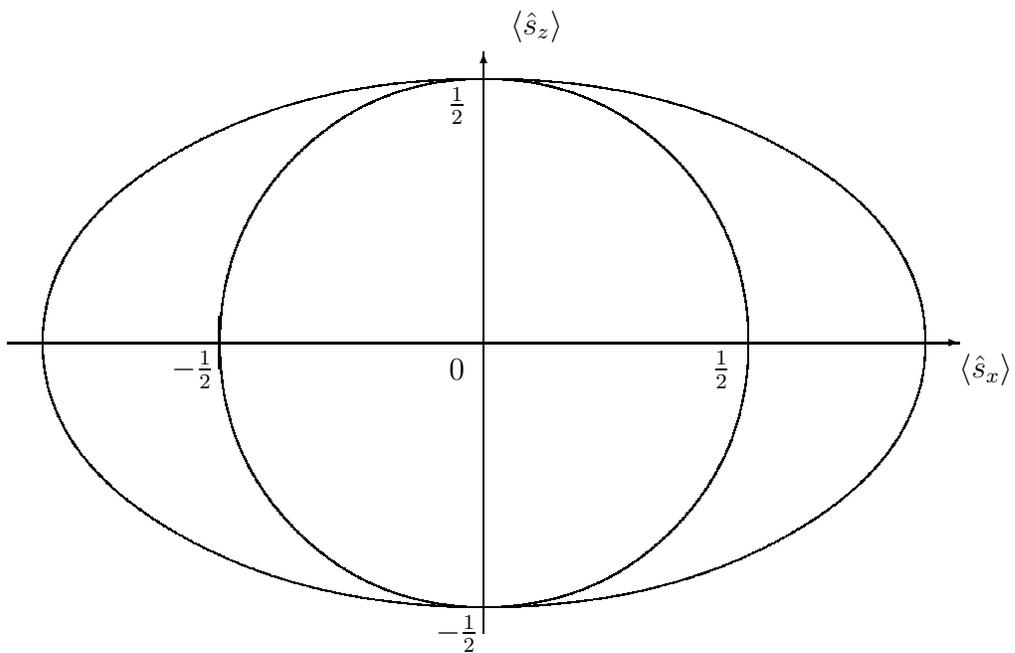

At this point, a specific example should help to illustrate the
case of negative probabilities after the measurement.
If the system concerned is a spin 1/2 system system described by the 
two orthogonal eigenstates of the $\hat{s}_z$ component,  
$\mid +Z \rangle$ and $\mid -Z \rangle$, then all physical properties
can be described in terms of the operators of the spin components,
\begin{eqnarray}
\hat{s}_x &=& \frac{1}{2}\hspace{0.3cm}\left(\mid +Z \rangle \langle -Z \mid 
                             + \mid -Z \rangle \langle +Z \mid\right)
\nonumber \\ 
\hat{s}_y &=&-\frac{i}{2}\left(\mid +Z \rangle \langle -Z \mid 
                             - \mid -Z \rangle \langle +Z \mid\right)
\nonumber \\
\hat{s}_z &=& \frac{1}{2}\hspace{0.3cm}\left(\mid +Z \rangle \langle +Z \mid 
                             - \mid -Z \rangle \langle -Z \mid\right).
\end{eqnarray}
If the initial state is given by the eigenstate of $\hat{s}_x$ with the
eigenvalue $+1/2$, $\mid +X \rangle$, then the initial density matrix reads
\begin{eqnarray}
\hat{\rho}_i &=& 
     \frac{1}{2} \bigg ( \mid +Z \rangle \langle +Z \mid
                      + \mid -Z \rangle \langle -Z \mid 
\nonumber \\ &&
                      + \mid +Z \rangle \langle -Z \mid
                      + \mid -Z \rangle \langle +Z \mid 
\bigg). 
\end{eqnarray}
The probability $p(\bar{s}_z)$ of obtaining a measurement result
of $\bar{s_z}\pm \delta\! s_z$ can be determined according to 
equation (\ref{eq:distrib}) using the corresponding 
generalized measurement operator. It reads
\begin{eqnarray}
p(\bar{s}_z) &=& \frac{1}{\sqrt{2\pi \delta\! s_z^2}}
\left(\frac{1}{2}\exp\left(-\frac{(\bar{s}_z-1/2)^2}{2\delta\! s_z^2}\right)
     +\frac{1}{2}\exp\left(-\frac{(\bar{s}_z+1/2)^2}{2\delta\! s_z^2}\right)
\right)
\nonumber \\
&=& \frac{1}{\sqrt{2\pi \delta\! s_z^2}}
             \exp\left(-\frac{\bar{s}_z^2+1/4}{2\delta\! s_z^2}\right)
             \cosh \left(\frac{\bar{s}_z}{2\delta\! s_z^2}\right).
\end{eqnarray}
Spin quantization clearly emerges in this probability
distribution if $\delta\! s_z$ is smaller than $1/2$. 
The noise free part of the measurement changes the density matrix to
\begin{eqnarray}
\hat{\rho}_m(\bar{s}_z) &=& 
     \frac{1}{2 \cosh \left(\frac{\bar{s}_z}{2\delta\! s_z^2}\right)}
\nonumber \\
&\times& 
\bigg(  \exp\left(\frac{\bar{s}_z}{2\delta\! s_z^2}\right)
                                 \mid +Z \rangle \langle +Z \mid
      +\exp\left(-\frac{\bar{s}_z}{2\delta\! s_z^2}\right) 
                                 \mid -Z \rangle \langle -Z \mid 
\nonumber \\ &&  
      +\exp\left(\frac{1}{8\delta\! s_z^2}\right)
                                 \mid +Z \rangle \langle -Z \mid
      +\exp\left(\frac{1}{8\delta\! s_z^2}\right) 
                                 \mid -Z \rangle \langle +Z \mid \bigg). 
\end{eqnarray}
This density matrix violates condition (\ref{eq:condition}) and
predicts negative probabilities for several spin directions. 
The negative probabilities can be illustrated by the expectation
values of the spin components,
\begin{eqnarray}
\langle \hat{s}_x \rangle_m &=& 
        \frac{\exp\left(\frac{1}{8\delta\! s_z^2}\right)}
             {2 \cosh \left(\frac{\bar{s}_z}{2\delta\! s_z^2}\right)}
\nonumber \\ &=& \frac{1}{2}
 \exp\left(\frac{1}{8\delta\! s_z^2}\right)
       \sqrt{1-\tanh^2 \left(\frac{\bar{s}_z}{2\delta\! s_z^2}\right)}
\nonumber \\
\langle \hat{s}_y \rangle_m &=& 0
\nonumber \\
\langle \hat{s}_z \rangle_m &=& \frac{1}{2}
        \tanh \left(\frac{\bar{s}_z}{2\delta\! s_z^2}\right).
\end{eqnarray}
Figure \ref{bloch} shows the expectation values in the xz plane for 
a measurement uncertainty of $\delta\! s_z^2 = 1/4$.
Except at $\langle \hat{s}_x\rangle=0$, the length of the average spin
vector is larger than $1/2$, indicating negative probabilities.
In particular, a result of $\bar{s}_z=0$ increases the expectation
value of $1/2$ in $\hat{s}_x$ to $1/2 \times \exp (1/(8\delta\! s_z^2))$.
This corresponds to a probability greater than one for $\mid +X \rangle$
and a corresponding negative probability for $\mid -X \rangle$.
Note that this result seems to be related to the observation of spin 
components larger than the permitted eigenvalue limit reported 
in \cite{Aha88}, even though actual measurements of the spin component
$\hat{s}_x$ are not considered in the present context. 

Of course, negative probabilities cannot be observed in a measurement.
They represent a statistical tool which is connected with the unavoidable
presence of quantum noise.
In order to recover the final density matrix 
$\hat{\rho}_f (\bar{s}_z)$
after the measurement, quantum noise must be added. 
This reduces the coherence to normal levels, resulting in the pure
state density matrix 
\begin{eqnarray}
\hat{\rho}_f (\bar{s}_z) &=& 
     \frac{1}{2 \cosh \left(\frac{\bar{s}_z}{2\delta\! s_z^2}\right)}
\nonumber \\
&\times& 
\bigg( \exp\left(\frac{\bar{s}_z}{2\delta\! s_z^2}\right)
                                 \mid +Z \rangle \langle +Z \mid
      +\exp\left(-\frac{\bar{s}_z}{2\delta\! s_z^2}\right) 
                                 \mid -Z \rangle \langle -Z \mid 
\nonumber \\ && \hspace{4cm} 
      + \mid +Z \rangle \langle -Z \mid
      + \mid -Z \rangle \langle +Z \mid \bigg). 
\end{eqnarray}
The expectation values given by this density matrix read
\begin{eqnarray}
\langle \hat{s}_x \rangle_f &=& 
         \frac{1}{2}
       \sqrt{1-\tanh^2 \left(\frac{\bar{s}_z}{2\delta\! s_z^2}\right)}
\nonumber \\
\langle \hat{s}_y \rangle_f &=& 0
\nonumber \\
\langle \hat{s}_z \rangle_f &=& \frac{1}{2}
        \tanh \left(\frac{\bar{s}_z}{2\delta\! s_z^2}\right),
\end{eqnarray}
as shown in figure \ref{bloch}.
In particular, the expectation value of $\hat{s}_x$ for
$\bar{s}_z=0$ is reduced to $1/2$.  
Thus, quantum noise is necessary in order to ``cover up'' any negative 
probabilities and any excessive expectation values arising in the noise 
free formalism.

\subsection{Interpretation of negative probabilities}
The example above shows that negative probabilities can compensate
the decoherence caused by the noisy measurement interaction. It could 
therefore be said that
negative probabilities represent non-classical information
beyond the limits of uncertainty. Indeed, the density matrix 
$\hat{\rho}_m$ may be ``purer'' than a pure state. The trace of the square
of a pure state density matrix is one. For a mixed state, it is smaller
than one. For $\hat{\rho}_m$, however, it may indeed be greater than one.
Thus, with the purity $P_m$ of $\hat{\rho}_m$ defined as
\begin{equation}
P_m=tr\{\hat{\rho}^2_m\},
\end{equation} 
the density matrix $\hat{\rho}_m(\bar{s}_z=0)$ given in the example of
section \ref{sec:example} above has a purity of 
\begin{equation}
P_m(\bar{s}_z=0) = \frac{1}{2}+
    \frac{1}{2}\exp\left(\frac{1}{4\delta\! s_z^2}\right),
\end{equation}
which is greater than one in all cases. Note that there is no upper
limit to $P_m(0)$. However, its increase does depend on the 
decrease in the likelihood of observing $\bar{s}_z=0$.  

Note that the ``super purity'' of the noise free density matrix
$\hat{\rho}_m$ corresponds to the classical notion that 
any new information gained about a physical system should reduce 
the uncertainty of the state. Therefore, obtaining information about
any state has to increase the purity of this state. 
Negative probabilities allow such an increase in purity even for a 
pure state. While ``super pure'' states are of course unphysical by 
themselves, they can be used to provide a local interpretation of
entanglement. In particular, entangled states can always be interpreted 
as a mixture of ``super pure'' product states. Classical probability 
theory then explains, why entanglement cannot be utilized to instantaneously
transfer information without physical interaction. Thus, the generalized
measurement operator may explain the physical nature of entanglement in
a far more intuitive way than the conventional formalism.

The change in the density matrix caused by a quantum measurement 
can now be compared with the changes caused by information obtained
about a classical probability distribution.  
A classical noise free measurement can only change the statistics of 
physical properties if the corresponding properties are correlated with
the measured variable. In the example given in section \ref{sec:example}, 
however, the statistics of $\hat{s}_x$ are changed by the measurement
even though the initial state is an eigenstate of $\hat{s}_x$. Classically,
a well defined variable cannot be correlated with any other variable.
By introducing negative probabilities, however, this situation is changed.
In the case above, the measurement reduces the expectation value of
$\hat{s}_x$ to a very low value if a value of $\bar{s}_z$ close
to the quantized values of $\pm 1/2$ is observed. On the other hand, 
negative probabilities appear if $\bar{s}_z$ is close to the average 
between the two quantized values. The original pure state is retained
if one averages over all measurement results, as shown by equation
(\ref{eq:decomp}). Thus, there is a statistical correlation between 
the spin component $\hat{s}_x$ and the spin component $\hat{s}_z$
which may be expressed by averages over the measurement results
$\bar{s}_z$ and the expectation values $\langle \hat{s}_x \rangle$
after the measurement as
\begin{equation}
\label{eq:corr}
\overline{\bar{s}_z^2 \langle \hat{s}_x \rangle}
-\overline{\bar{s}_z^2}\hspace{0.3cm} \overline{\langle \hat{s}_x \rangle} 
= - \frac{1}{4} \overline{\langle \hat{s}_x \rangle}  
.
\end{equation}
In words, the measurement of a quantized value of $\hat{s}_z$ is correlated
with the non-vanishing possibility of a negative value of $\hat{s}_x$,
while the measurement of a value of $\hat{s}_z$ between the quantized values
is correlated with a negative probability for  $\hat{s}_x<0$, 
indicating that such negative values are ``more than impossible''.

\subsection{Nonclassical correlations as fundamental operator properties}
Usually, entanglement is analyzed in terms of operator properties.
In particular, nonclassical features of spin-1/2 statistics can often
be traced to the anti commutation of the spin components. Using the 
generalized projection operator $\hat{P}_{\delta\!A}(\bar{A})$, it is
possible to derive an analytical expression for the correlation of 
$\bar{s}_z^2$ and $\langle \hat{s}_x \rangle$ given in equation  
(\ref{eq:corr}). In general, the correlation between the squared measurement
result $\bar{A}^2$ and a variable $\hat{B}$ after the measurement is 
given by
\begin{eqnarray}
C(\bar{A}^2;\langle\hat{B}\rangle)&=&
\overline{\bar{A}^2 \langle \hat{B} \rangle}
-\overline{\bar{A}^2}\hspace{0.3cm} \overline{\langle \hat{B} \rangle} 
\nonumber \\
&=& \int d\! \bar{A} \; p(\bar{A}) Tr\left\{\rho_f(\bar{A}) \hat{B}\right\} 
\bar{A}^2 \nonumber \\
&& - Tr\left\{\rho_f(\mbox{total}) \hat{B}\right\}\; 
\left(\delta\!A^2 + Tr\left\{\rho_f(\mbox{total}) \hat{A}^2\right\}\right).
\end{eqnarray}
By solving the integral over $\bar{A}$, the correlation may be expressed
entirely in terms of expectation values of the final state density
matrix $\hat{\rho}_f(\mbox{total})$. It reads
\begin{equation}
C(\bar{A}^2;\langle\hat{B}\rangle) = 
\frac{1}{4} \langle \hat{A}^2\hat{B} + 2 \hat{A}\hat{B}\hat{A} + 
\hat{B}\hat{A}^2\rangle_f(\mbox{total}) 
- \langle \hat{A}^2 \rangle _f(\mbox{total})
  \langle \hat{B} \rangle _f(\mbox{total}).  
\end{equation}
Since $\hat{B}$ does not necessarily commute with $\hat{A}$, this correlation
can be nonzero even if the system is in an eigenstate of $\hat{A}$. In the
case of the spin-1/2 system discussed above, the anti commutation of $\hat{s}_x$
and $\hat{s}_z$ is responsible for the result that
\begin{equation}
C(\bar{s}_z^2;\langle\hat{s}_x\rangle) =  
- \langle \hat{s}_z^2 \rangle _f(\mbox{total})
  \langle \hat{s}_x \rangle _f(\mbox{total}).  
\end{equation}

In the case of $\delta\!s_z\to \infty$, or when removing the noise
effect from the measurement, the final density matrix can be replaced with
the initial state. It is then possible to obtain nonzero correlations between
$\hat{s}_x$ and $\hat{s}_z$ even for the eigenstates of $\hat{s}_x$.
This result suggests that the physically relevant value of $\hat{s}_x$ 
is not even well defined for eigenstates of $\hat{s}_x$! 
In the words of the famous EPR paper
\cite{EPR}, the fact that the outcome of an $\hat{s}_x$ measurement
can be predicted with certainty does not mean that there exists an element
of reality corresponding to this {\it potential} measurement result, unless 
the measurement is {\it actually} performed. The $\hat{s}_x$-fluctuations of 
an eigenstate of $\hat{s}_x$ are revealed in the noise induced changes
of $\hat{s}_x$ if any property other than $\hat{s}_x$ interacts with the
environment.

\subsection{Negative probabilities and quantization}
One of the fundamental consequences of quantum mechanics is the
replacement of continuous classical variables with discrete quantum
numbers. In particular, the components of angular momentum 
have eigenvalues equal to multiples of $\hbar$ and the
energies of harmonic oscillators or wave modes have eigenvalues equal
to multiples of $\hbar\omega$. One of the strangest features of quantum
mechanics is the inconsistency this introduces into classical arguments.
For example, it should be necessary to conclude that, if $\hat{s}_x$
is equal to $\pm 1/2$ and $\hat{s}_y$ is equal to $\pm 1/2$, $\hat{s}_x
+ \hat{s}_y$ should be equal to zero or $\pm 1$. However, the eigenvalues
of   $\hat{s}_x + \hat{s}_y$ are $\pm 1/\sqrt{2}$. It is this contradiction
of classical arguments based on quantized values which is exploited in the
formulation of Bell's inequalities \cite{Bel64}. 
Usually, one tries to escape the
dilemma by arguing that the classical meaning of the quantized observable
is lost completely. Alternatively, however, one could assume that eigenvalues
do {\it not} represent all physical values of the observable. 
The observation of
quantized values in precise measurements could instead be explained as 
a fundamental statistical effect based on the presence of negative 
probabilities and quantum correlations. 

In the example above, the eigenvalues of $\hat{s}_z$ are $\pm1$. However,
there is a correlation between $\bar{s}_z^2$ and $\hat{s}_x$ which suggests
some measure of physical reality for fluctuations in $\bar{s}_z^2$, especially
for the possibility of non-quantized values near $\bar{s}_z=0$.
In the case of light field quantization, the same principles can be applied 
to quantum nondemolition measurements of photon number \cite{Hof00}, revealing
high phase coherence at half integer photon numbers and low phase coherence
at integer photon numbers.
The measurement process can now be analyzed in far greater depth.
If separate quantum states are not resolved, negative probabilities 
and quantum correlations are hidden by the remaining uncertainty in the
observed variable. If quantization is resolved, quantum correlations
modify the statistics of all variables that do not commute with the 
observed variable. Such correlations can only be interpreted in terms
of negative probabilities. However, quantum noise ``covers up'' the 
negative probabilities. Nevertheless, negative probabilities can be
observed indirectly in nonclassical correlations. 

Recently, the question of what truly characterizes the differences between 
classical physics and quantum physics has been raised in a new context
regarding the potential of quantum computers \cite{Fit00,Bra99}. It seems
that on the quite technical level represented e.g. by NMR quantum computation,
the statistical relationships between those operator variables actually
utilized are far more important than observable independent concepts such as
entanglement. Possibly, contemporary quantum theory has paid too little
attention to the observable properties of quantum systems. By interpreting
quantum mechanics in terms of nonclassically correlated observables, a 
smooth transition between the classical regime and the quantum regime is
possible and the problem of suddenly having to change the vocabulary from 
physical properties to Hilbert spaces can be avoided. Quantum properties
can then be explored within a framework similar to that of classical physics,
with the main quantum correction originating from the nonclassical correlations
possible due to the appearance of negative probabilities. It should then be 
possible to identify the correlations required for quantum computation and
other applications of quantization effects.

\section{Conclusions}
In conclusion, a physical interpretation of the measurement process
based on a separation of information and noise is possible.
This separation corresponds exactly to the classical notion of a
reality unchanged by the measurement interaction. 
However, negative probabilities appear in the measurement decomposition of
the initial density matrix $\hat{\rho}_i$ into the conditioned density 
matrices $\hat{\rho}_m$ given by equation (\ref{eq:decomp}). 
These negative probabilities represent a type of nonclassical information 
only available in quantum mechanical systems. They are responsible for the
failure of measurement independent concepts of local reality such as
the one proposed in the EPR paper\cite{EPR}, and 
it is likely that this type of nonclassical information is also 
responsible for the advantages of quantum computing as compared 
with classical computing.  
Moreover, the negative probabilities are directly related to quantization
itself, since they arise from correlations which distinguish between the
observation of quantized values and the observation of values between
two quantized eigenvalues of the observable. 

These results clearly show that there is much more to quantum 
measurements than the observation of eigenvalues. Possibly, the main
interpretational problem in quantum measurement theory is the 
assumption that physical variables should be restricted to their 
eigenvalues. However, negative
probabilities and the real physical consequences of measuring a value
between two eigenvalues seem to indicate that the effective physical 
properties of a variable are not restricted to eigenvalues
only. Instead, some measure of physical reality should be attributed to the
continuum of values between and even beyond the eigenvalues.
In particular, the off-diagonal matrix elements of the density matrix
can be associated with the average of the two associated eigenvalues,
even if this average does not correspond to any actual eigenvalue.
Quantum coherence can then be understood in terms of nonclassical correlations
between the physical properties of a system.
Thus the generalized measurement postulate represents an opportunity for
developing new interpretational concepts in quantum theory which may 
allow us to improve our intuitive understanding of the physical nature
of quantum effects.

\section*{Acknowledgements}
The Author would like to acknowledge support from the Japanese Society for
the Promotion of Science, JSPS.

\newpage
%


\end{document}